\documentclass{aa}
\usepackage{amsmath,graphics,amssymb}
\usepackage{times}

\allowdisplaybreaks

\newfont{\eufont}{eufm10}
\def\eu #1{\mbox{\eufont #1}}

\newcommand\refi{}

\newcommand{\dmdt}{\dot{\text{\eu M}}}
\newcommand{\hm}{{\text{\eu M}}}

\newcommand{\io}[1]{{#1}_{\mathrm{i}}}
\newcommand{\zav}[1]{\left(#1\right)}

\newcommand{\xa}{\io{\eu Y}}

\newcommand\jcb{}

\begin{document}

\title{On multicomponent effects in stellar winds of stars at extremely
low metallicity}

\author{J.  Krti\v{c}ka\inst{1,2,3} \and S. P. Owocki\inst{1,4}
\and J. Kub\'at\inst{3}
\and R. K. Galloway\inst{1}
\and J. C. Brown\inst{1}}
\authorrunning{J.  Krti\v{c}ka et al.}

\offprints{J. Krti\v{c}ka, \email{jiri@astro.gla.ac.uk}}

\institute{Department of Physics and Astronomy, University of Glasgow, 
           Glasgow G12 8QQ, UK
           \and
           \'Ustav teoretick\'e fyziky a astrofyziky P\v{r}F MU,
            CZ-611 37 Brno, Czech Republic
           \and
		   Astronomick\'y \'ustav, Akademie v\v{e}d \v{C}esk\'e
           republiky, CZ-251 65 Ond\v{r}ejov, Czech Republic
           \and
           Bartol Research Institute, University of Delaware, Newark,
           DE 19716, USA}

\date{Received 29 November 2002 / Accepted 17 February 2003}

\abstract{
We calculate multicomponent line-driven wind models of stars 
at extremely
low metallicity suitable for massive first generation stars. 
For most of the models we find that the multicomponent wind nature is not important for either
wind dynamics or for wind temperature stratification. However,
for stars with the lowest metallicities we find that multicomponent effects
influence the wind structure. These effects range from pure heating to possible
fallback of the nonabsorbing wind component.
We present a simple formula for the calculation of metallicity for
which the multicomponent effects become important. 
We show that the importance of the multicomponent nature of winds of low
metallicity stars is characterised not only by the low density of driving ions, but also by
lower mass-loss rate.

}

\maketitle

\section{Introduction}

First generation stars are a textbook example of the importance of initial
metallicity for stellar structure and evolution.
During the gravitational collapse of extremely-low metallicity clouds
the formation of massive stars was possibly much more
favored than in the present time (cf. Bromm et al. \cite{bcl}, Nakamura \&
Umemura \cite{naum}) and the evolution of such  massive stars was strikingly
different from standard stellar evolution (Siess et al. \cite{evol}).
Similarly, due to their enormous luminosity and minute
abundance of elements heavier than helium, their line-driven
winds substantially differed from those of present day stars.

The first models of line-driven stellar winds suitable for very massive
first generation
stars were calculated by Kudritzki (\cite{kudmet}). He concluded that these
stars must be near to the Eddington limit in order to possess line-driven
winds. However, in the case of stars at such extremely low metallicity
other effects may become important. It is well known that radiative driven
stellar winds have a multicomponent nature (cf. Castor et al. 
\cite{cakvice}), the
reason for which is relatively straightforward. The radiative
force accelerating a stellar wind is distributed unevenly over individual ions.
Whereas minor absorbing elements, such as C, N, O or Fe, obtain momentum from
the stellar radiation field, hydrogen and and helium are only marginally accelerated by the
radiation. 
However, these nonabsorbing components are
accelerated by friction with the absorbing components.  This momentum transfer
between low-density absorbing and high-density nonabsorbing components is
specially important for low-density stellar winds (Springmann \&
Pauldrach \cite{treni}, Krti\v{c}ka \& Kub\' at \cite{kkii}, hereafter KKII).
These stellar winds can be heated by the friction between
components and, for extremely low wind density, even hydrogen fallback may
occur. Low density stellar winds may also be subject to so called runaway
instability (see Owocki \& Puls
\cite{op}, Krti\v{c}ka \& Kub\' at \cite{kkiii}). This instability occurs when
the velocity difference between absorbing and nonabsorbing components is
comparable to the averaged sound speed
(Krti\v{c}ka \& Kub\' at \cite{kkii}, Eq. (45)).
Finally, low-density stellar winds are
heated by the so-called Gayley-Owocki heating (Gayley \& Owocki 
\cite{go}, hereafter
GO). This heating/cooling is caused by the dependence of the 
radiative force on the
velocity via the Doppler effect.

Because these multicomponent effects usually
occur mainly in the outer part of the wind
downstream from the critical point they do not affect the mass-loss 
rate but they
influence the wind temperature and outflow velocity.
In this paper we discuss these multicomponent effects for stars at 
extremely low
metallicity.

\section{Domain of importance of multicomponent effects}

In order to assess the importance of multicomponent effects we use a simple
formula for the velocity difference between absorbing and 
nonabsorbing components.
Similarly to Springmann \& Pauldrach (\cite{treni}) and Owocki \& Puls
(\cite{op})
we start from the equation of motion of nonabsorbing component
for a two-component wind.
Neglecting the
gravitational acceleration, electrical force and gas pressure term, 
this equation
takes the form (see KKII)
\begin{equation}
\label{papro}
v_{r\mathrm{p}}\frac{\mathrm{d}v_{r\mathrm{p}}}{\mathrm{d}r}=
\frac{\rho_{\mathrm{i}}}{m_\mathrm{p} m_{\mathrm{i}}}\frac{4\pi 
q_{\mathrm{p}}^2
q_{\mathrm{i}}^2}{kT}\ln\Lambda G(x_{\mathrm{pi}}),
\end{equation}
where $v_{r\mathrm{p}}$ is the velocity of the nonabsorbing component.
For the calculation of frictional acceleration on the right-hand side of
Eq.(\ref{papro}) we assume that the temperatures of
all components
are nearly
equal ($T_\mathrm{p}\approx T_{\mathrm{i}}\approx T$), and $\rho_{\mathrm{i}}$
is the
mass density of
the absorbing component, $m_\mathrm{p}$, $m_{\mathrm{i}}$, $q_{\mathrm{p}}$ and
$q_{\mathrm{i}}$ are the particle masses and charges of nonabsorbing 
and absorbing
components, and $\ln\Lambda$ is the Coulomb logarithm. The Chandrasekhar
function
$G(x_{\mathrm{pi}})$
{\refi is defined in terms of the error function $\mathrm{erf}(x)$  
(cf. Burgers \cite{burgers})}
\begin{align}
G(x_{\mathrm{pi}})&=\frac{1}{2
x_{\mathrm{pi}}^2}\zav{\mathrm{erf}(x_{\mathrm{pi}})-\frac{2x_{\mathrm{pi}}}{\sqrt\pi}\exp\zav{-x_{\mathrm{pi}}^2}},\\
x_{\mathrm{pi}}&\approx\frac{v_{r\mathrm{i}}-v_{r\mathrm{p}}}
{\sqrt{\frac{2kT}{m_{\mathrm{p}}}}}.
\end{align}
{\refi The Chandrasekhar
function can be approximated for $x_{\mathrm{pi}}<1$} by
\begin{equation}
G(x_{\mathrm{pi}})\approx\frac{2 x_{\mathrm{pi}}}{3\sqrt{\pi}}. 
\end{equation}
Using a $\beta$-velocity law 
$v_{r\mathrm{p}}=v_{\infty}(1-\frac{R_*}{r})$ 
(with $\beta=1$, {\jcb which is near to the mean observed value, cf. Puls et al.
\cite{puls}})
for the non-absorbing component we can approximate
$\frac{\mathrm{d}v_{r\mathrm{p}}}{\mathrm{d}r}\approx 
v_{\infty}\frac{R_*}{r^2}$.
The continuity equation can be used to calculate the density of absorbing ions
\begin{equation}
\rho_{\mathrm{i}}\approx \zav{Z\over {Z_\odot}} \xa\rho_{\mathrm{p}}
 \approx \frac{1}{4\pi r^2 v_{r\mathrm{p}}}   \zav{Z\over {Z_\odot}}  \xa\dmdt\, ,
\end{equation}
where $v_{\infty}$ and $R_*$ are the terminal velocity and stellar 
radius, $\xa$ is
the 
mass density ratio of absorbing and nonabsorbing ions
{\refi in the solar photosphere ($\xa=0.0127$, this value 
 corresponds to the solar ratio of sum of densities of
 C, N, O, Fe to the density of bulk plasma),}
$Z/Z_\odot$ is the metallicity
{\jcb (number density of absorbing ions relative to hydrogen)
in the stellar atmosphere} relative to the solar value,
and $\dmdt$ is mass loss rate.
Solving the momentum equation
\eqref{papro}
for the velocity difference
we obtain
\begin{equation}
\label{xpi}
\frac{v_{r\mathrm{i}}-v_{r\mathrm{p}}}
{\sqrt{\frac{2kT}{m_{\mathrm{p}}}}}
\approx v_{r\mathrm{p}}^2
\frac{v_{\infty}R_*}{(Z/Z_\odot) \dmdt}
\frac{3\sqrt{\pi}m_{\mathrm{p}}m_{\mathrm{i}}kT}{2\xa 
q_{\mathrm{p}}^2q_{\mathrm{i}}^2\ln\Lambda}.
\end{equation}
Apparently, as the wind is accelerated the velocity difference increases. Thus,
multicomponent effects are important mainly in the outer part of the 
wind. If the
wind temperature is constant then the maximal velocity difference 
is attained
for maximal radial velocity, ie. for 
$v_{r\mathrm{p}}=v_{\infty}$. This
enables us to calculate the maximal velocity difference between wind 
components for
each model star.

\begin{figure}
\resizebox{\hsize}{!}{\input{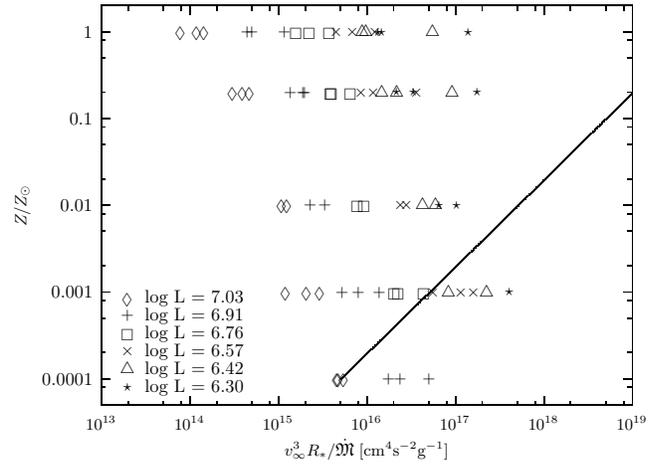}}
\caption[]{The domain of importance of multicomponent effects. Each model star
is represented by a point in this plot. The straight line (calculated using
Eq.(\ref{xpi}),
where we inserted $v_{r\mathrm{p}}=v_\infty$)
divides regions where multicomponent
effects are negligible from where they are
important. Winds of stars above this line have a low velocity 
difference between
wind components whereas for stars below the line multicomponent effects are
important. Different symbols denote different stellar
luminosity (given in solar
units).}
\label{vliv}
\end{figure}

The multicomponent effects are found to be important when the 
nondimensional velocity
difference is typically
\begin{displaymath}
\frac{v_{r\mathrm{i}}-v_{r\mathrm{p}}}{
\sqrt{\frac{2kT}{m_{\mathrm{p}}}}}\gtrsim 0.1
\end{displaymath}
(KKII).
In this case frictional heating influences the wind temperature. Note
that there is some kind of feedback because the higher temperature enhances the
multicomponent effects.
In the plot of metallicity versus the quantity $ v_{\infty}^3R_*/\dmdt$
(see Fig.\ref{vliv})
we shall use a straight line to divide regions where multicomponent 
{\refi effects}
are
important/unimportant. Stars
above this line have a low relative velocity difference between wind 
components (see
Eq.(\ref{xpi})) whereas stars below this line have higher  relative velocity
difference and thus multicomponent effects are important for these stars.
{\refi The dividing line (for which $x_\mathrm{pi}=0.1$)
was plotted with assumed average wind temperature
$T=25\,000\,$K,
ionic charges $q_\mathrm{p}=0.85 e$, $q_\mathrm{i}=4.0 e$
(where $e$ is
the electronic charge) and the mean mass of absorbing ions
corresponding to carbon.
}

We added each star for which Kudritzki (\cite{kudmet}) calculated a wind
model into Fig.\ref{vliv}. 
{\refi We used the same wind parameters 
(ie.~wind temperature, ionic charges and mass of absorbing ions)  as in 
the
previous
paragraph for the plot of dividing line.}
There are several stars for which the multicomponent effects are
important. For stars with highest luminosity only
the models with lowest metallicity may suffer from the multicomponent effect.
The lower the luminosity,
{\jcb the lower the mass loss rate and}
the higher is the  metallicity for which 
multicomponent
effect becomes important. Moreover,
for stars with given basic parameters (ie. mass, radii and effective 
temperature) we
are able to find a value of metallicity for which the multicomponent effects
influence the wind structure. This conclusion is especially important for low
metallicity stars.

{\jcb
For normal solar-type ion abundance with 
$(Z/Z_\odot)\approx1$
Eq.
\eqref{xpi}
indicates that ion runaway should occur only for
very low mass loss rates, i.e. $\dmdt \sim 10^{-11} M_\odot $/yr,
and so is relevant only for relatively low luminosity stars.
But for stars with very low metallicity, ion runaway could become
important for quite luminous stars. 
Consider, for example, the canonical CAK mass loss scaling
(see Pauldrach et al. \cite{ppk})
\begin{equation}
\label{dmdtreq}
{\dmdt} \approx {L \over c^2} 
{\alpha \over (1-\alpha) (1+\alpha)^{1/\alpha}}
 \left[ {{\bar Q \Gamma} \over 1-\Gamma} \right]^{-1+1/\alpha} \, ,
\end{equation}
where $\Gamma$ is the Eddington parameter, and
${\bar Q} \approx 2000\, Z/Z_\odot$ (Gayley \cite{gayley}).
Because
$\alpha<1$ (Abbott \cite{abpar}, 
Puls et al. \cite{rozdelfce}) this
generally means that lower metallicity enables larger multicomponent effects
not only due to the lower 
abundance of the driving ions
but also due to the
lower mass-loss rate
(see Eq.(\ref{xpi})).
The mass-loss rate formula
\eqref{dmdtreq}
can be rewritten using the 
scaled quantities 
${\dot M}_{-11} \equiv {\dmdt}/(10^{-11} \hm_\odot/\mathrm{yr})$,
$M_{2} \equiv \hm/10^2 \hm_\odot$, $L_{6} \equiv L/10^6 L_\odot$, and $\Gamma
\approx 0.26 L_6/M_2$.
For the canonical case of $\alpha=2/3$
and $\Gamma=0.5$,
this yields
\begin{equation}
\label{dmdtap}
{\dot M}_{-11} \approx
2.1 \times 10^5  { L_6^{3/2} \over M_2^{1/2}}
     \left( {Z \over Z_\odot} \right)^{1/2} \, .
\end{equation}
Similarly, Eq.(\ref{xpi}) implies that the maximal velocity difference scales as
\begin{equation}
\label{xpiap}
x_\mathrm{pi} \approx
1.2\, \frac{v_{8}^{3}R_{12}T_4}{{{\dot M}_{-11}} }\zav{Z \over Z_\odot}^{-1}
\end{equation}
where $v_{8}\equiv v_\infty/(10^8 \mathrm{cm}\,\mathrm{s}^{-1})$, 
$R_{12}=R_*/10^{12}\mathrm{cm}$, and $T_4\equiv T/10^4$K. Recalling that multicomponent effects are
important for $x_\mathrm{pi}\gtrsim 0.1$,
then inserting this value into
\eqref{xpiap}
we can obtain the minimal metallicity
$Z_{(1)}$ for
which one-component models are applicable.
By application of
\eqref{dmdtap} in \eqref{xpiap}
we find that
\begin{equation}
\zav{Z_{(1)} \over Z_\odot}
\approx
1.5\times 10^{-3}\,
\frac{v_{8}^2R_{12}^{2/3}T_4^{2/3}M_2^{1/3}}{L_6}.
\end{equation}
The latter equation can be rewritten in terms of basic stellar parameters 
using a canonical estimate of the wind terminal velocity
$v_\infty^2=\alpha/(1-\alpha)v_{\mathrm{esc}}^2$ (see
Castor et al. \cite{cak}, hereafter
CAK), where $v_{\mathrm{esc}}$
is the escape velocity. Thus, the lowest metallicity for which one-component wind models can be applied scales as}
\begin{equation}
\label{metjs}
\zav{Z_{(1)} \over Z_\odot}
\approx
4\times 10^{-3}\,
\frac{T_4^{2/3}M_2^{4/3}}{R_{12}^{1/3}L_6}.
\end{equation}
In any case, the general result from Eq.(\ref{metjs}) is that multicomponent
effects are likely
to play a role only in very low metallicity massive stars.

\section{Model description}

To test this scaling analysis we calculated multicomponent models of stars 
initially modelled by Kudritzki
(\cite{kudmet}) and looked for multicomponent effects.
In order to obtain correct models for massive stars at extremely low 
metallicity
and to improve the model convergence we used slightly different 
models from KKII. These changes are not important
  for the final results
since these models retain basics characteristics of the KKII models.

We assume
that a stellar wind consists of three components, namely absorbing 
driving
ions, passive nonabsorbing ions (hydrogen and helium, which mostly contribute
to the bulk wind density) and electrons.
For absorbing and nonabsorbing ions we solve the continuity equations, momentum
equations, and energy equations. The most important terms in the momentum
equations are the radiative and frictional acceleration, while the 
most important
terms in the energy equation are the heat exchange between wind components and
the GO heating. { Radiative force is calculated using the Sobolev approximation
(CAK)
with finite disk correction factor (Friend \& Abbott \cite{fa},
Pauldrach et al. \cite{ppk}).}
 Contrary to KKII, for the electrons we solve the
energy equation 
only{\refi. The} radiative heating/cooling term {\refi in this equation is}
calculated using
thermal balance of electrons (Kub\'at et al. \cite{kpp}).
{\jcb Stellar fluxes at the lower boundary of the wind were determined using
models calculated by a code of Kub\'at (\cite{kub}).}
Electron 
density and velocity are calculated from
the conditions of electrical neutrality and zero current. The
charge of the nonabsorbing component is calculated using ionization balance of
hydrogen and helium. We included helium  to obtain the correct
electron density. This proved to be crucially important in getting 
correct results
for stars near the Eddington limit.
Finally, we included new line-force multipliers introduced by Kudritzki
(\cite{kudmet}) both into the solved equations (with inclusion of the critical
point condition) and into the linearization matrix
{\jcb (see Krti\v{c}ka \cite{krt}).}

The output of these models
is the hydrodynamic structure of a three-component radiatively driven stellar
wind, i.e. density, velocity, temperature and charge of all components.
For a more detailed description of these models see KKII.

Originally, for the  calculation of the radiative force CAK 
introduced a parameter related to the thermal velocity, calculated at the stellar 
effective temperature. However, the radiative force
itself does not depend on the thermal velocity.
KKII
assumed an artificial dependency of this thermal
velocity parameter
on the local temperature instead of on the stellar effective temperature.
To come closer to models calculated by Kudritzki
(\cite{kudmet})
and to obtain a correct value of radiative force
we assumed that the thermal velocity parameter
used for the calculation of radiative force is constant.

The temperature dependence of the radiative force has, however,
an important impact on the calculated models.
If the radiative force does not depend significantly on the wind temperature (as
was consistently
assumed in this paper) then the velocity difference between wind components can
reach the sound speed and runaway instability occurs. On the other hand, if
the radiative force is decreasing with increasing temperature (as was 
assumed by KKII) then only heating occurs, and is
accompanied by the lowering of the terminal
velocity, or by the backfall of hydrogen for stars with extremely low-density
winds.

\section{Calculated models}

Generally, calculated models where multicomponent effects are important can be
divided into three groups according to the influence of these effects.
Each of this model groups
will
be discussed separately.

\subsection{Modified temperature structure}

For stellar winds with only mild maximal velocity difference
$(v_{r\mathrm{i}}-v_{r\mathrm{p}})/\sqrt{\frac{2kT}{m_{\mathrm{p}}}}\approx
0.1$ the only effect of the multicomponent flow is  a slightly 
modified temperature
structure compared to one-component models. Because the radiative force for
models in this paper
does not depend on the temperature, the outflow velocity 
is the same
as for the one-component models. This is a difference compared to 
KKII where for
this type of stars lowering of the outflow velocity occurs due to the lowering
of the radiative acceleration.

\begin{figure}
\resizebox{\hsize}{!}{\includegraphics{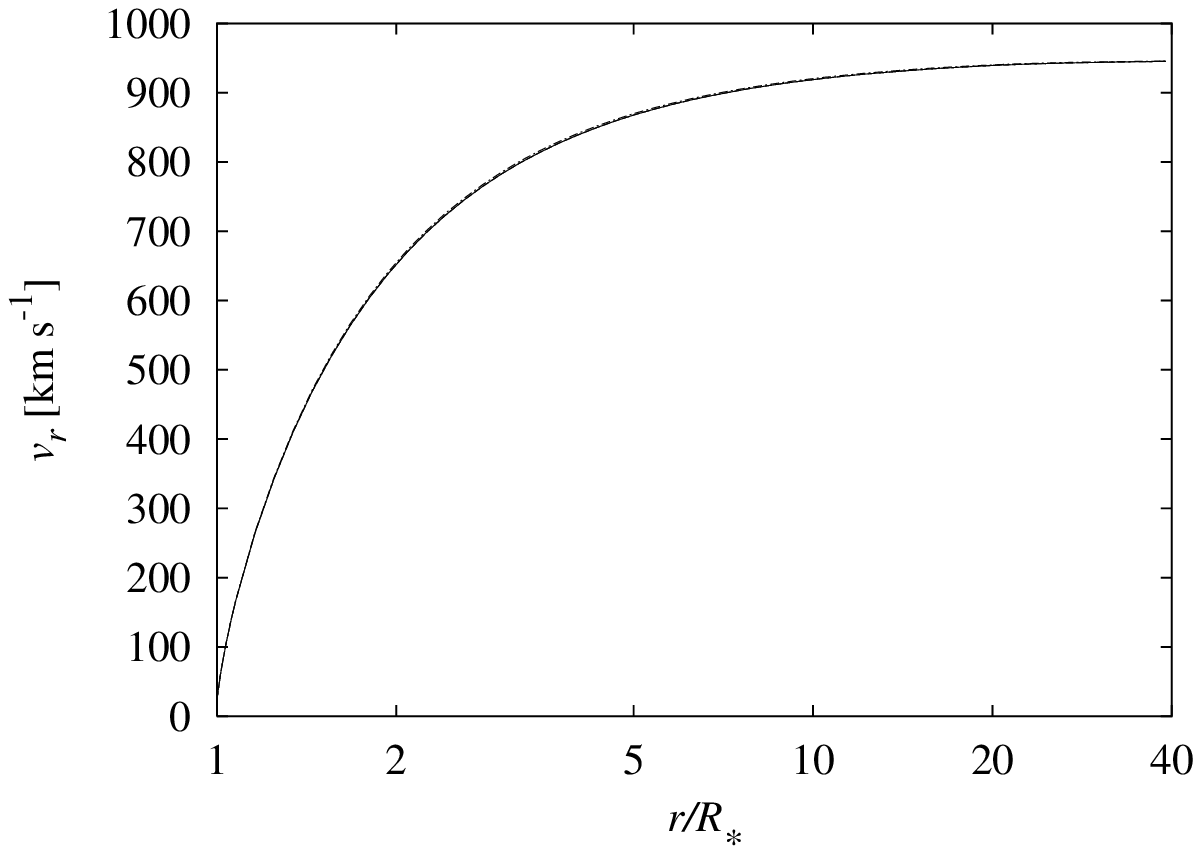}}
\resizebox{\hsize}{!}{\includegraphics{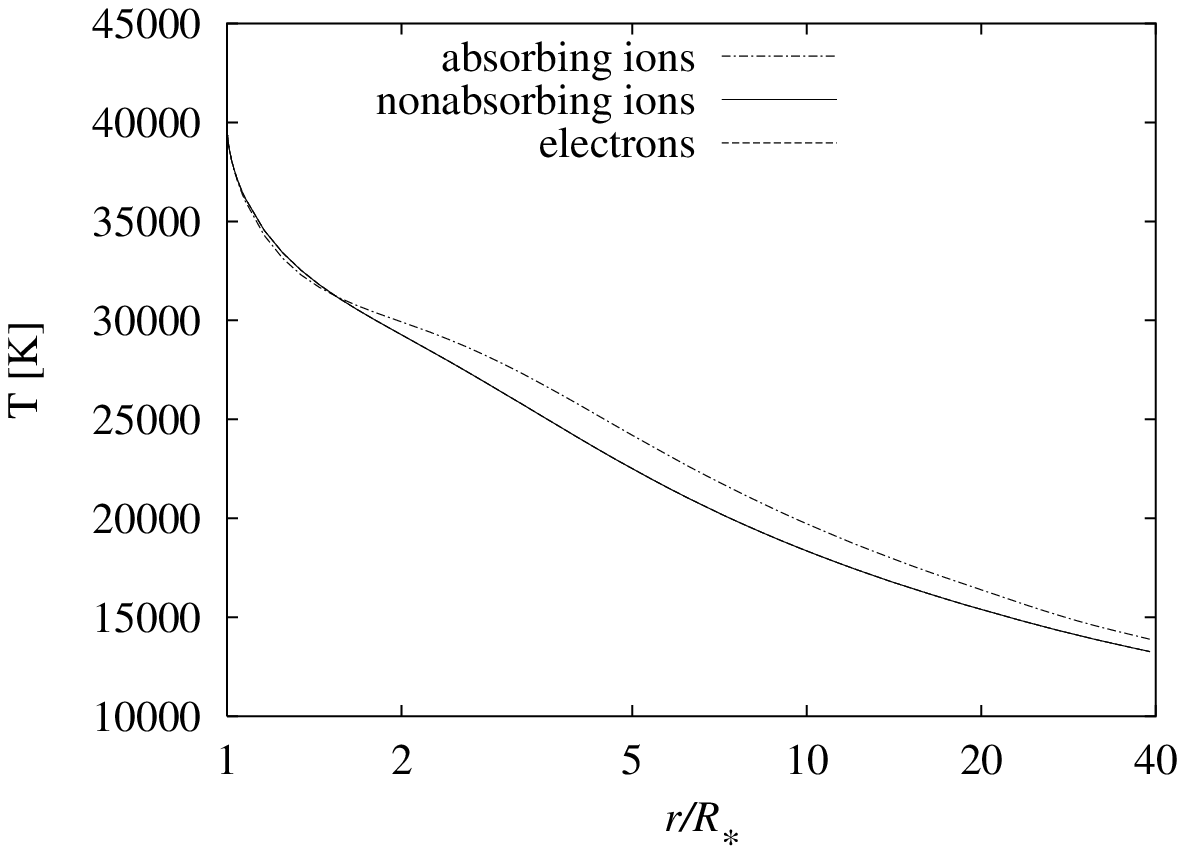}}
\caption[]{The calculated model of a star with $\log L/L_\odot=6.57$,
$T_\mathrm{eff}=50\,000\,\mathrm{K}$ and $Z/Z_\odot=0.001$. The stellar wind
temperature structure is modified by frictional and GO heating.
The temperature of absorbing ions is slightly higher than the temperature of
other wind components.
Note that velocities and densities of
nonabsorbing component and electrons are nearly the same. }
\label{mod}
\end{figure}

An example of such models is given in Fig.\ref{mod}. In this case the
temperature structure is affected by the frictional and GO heating. 
However, the
increase of the wind temperature is lower, compared to models of
main-sequence stars with similar velocity difference,
mainly due to the
{\refi higher wind density and consequently}
larger radiative cooling.   The wind
velocity is not changed compared to the one-component case due to the neglected
temperature dependence of the radiative force.

\subsection{Runaway instability in the outer region}

For stars with lower metallicity the wind density of absorbing component is
lower and, thus, the velocity difference between wind components is higher in
order to maintain common flow (see Eq.(\ref{xpi})). The lower the 
density, the higher the frictional
heating. However, this picture changes for extremely low-density
winds when the velocity difference is comparable to the sound speed,
$(v_{r\mathrm{i}}-v_{r\mathrm{p}})/\sqrt{\frac{2kT}{m_{\mathrm{p}}}}\approx 1$.
The wind is not
stable in this case for the ionic Abbott waves any more (see Owocki \& Puls
\cite{op}, Krti\v{c}ka \& Kub\' at \cite{kkiii}) and the runaway instability
occurs. This case did not occur in KKII because the temperature dependence of
radiative force prevented the velocity difference to attain higher values
comparable to the sound speed.

\begin{figure}
\resizebox{\hsize}{!}{\includegraphics{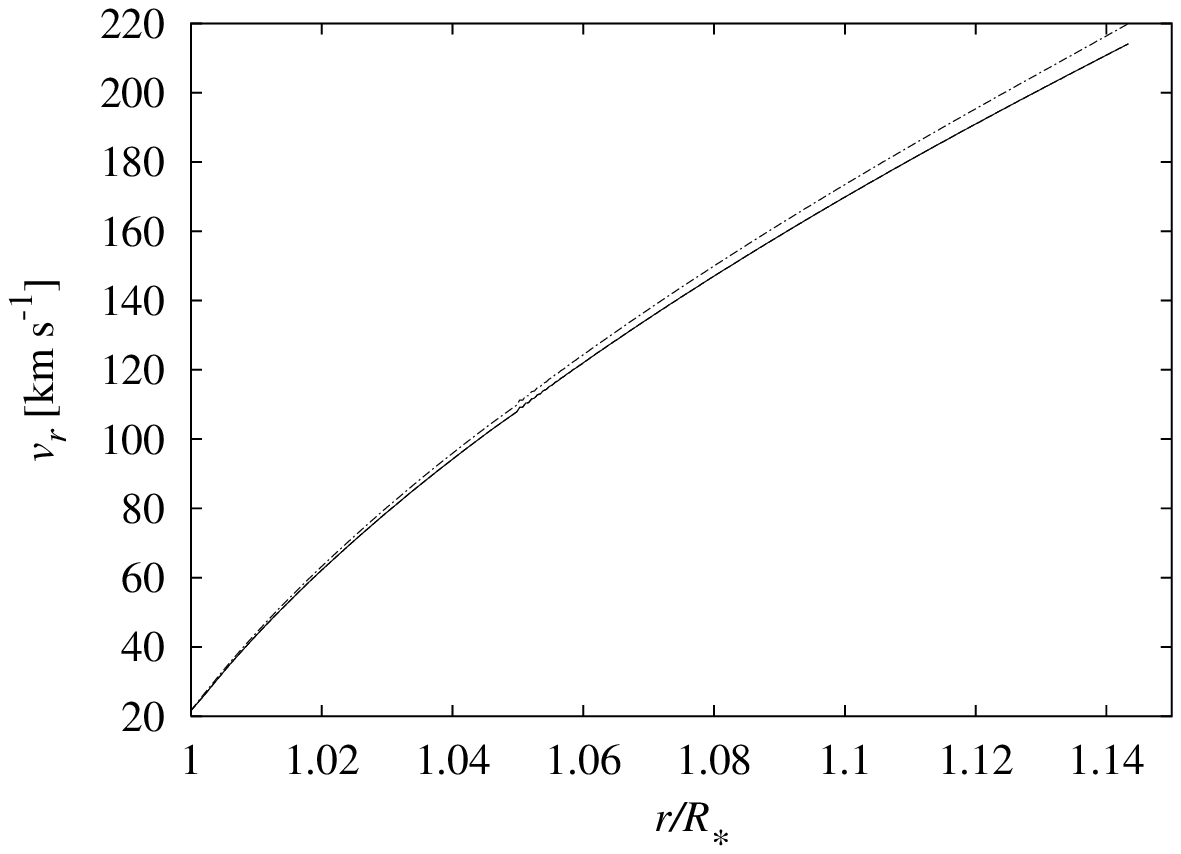}}
\resizebox{\hsize}{!}{\includegraphics{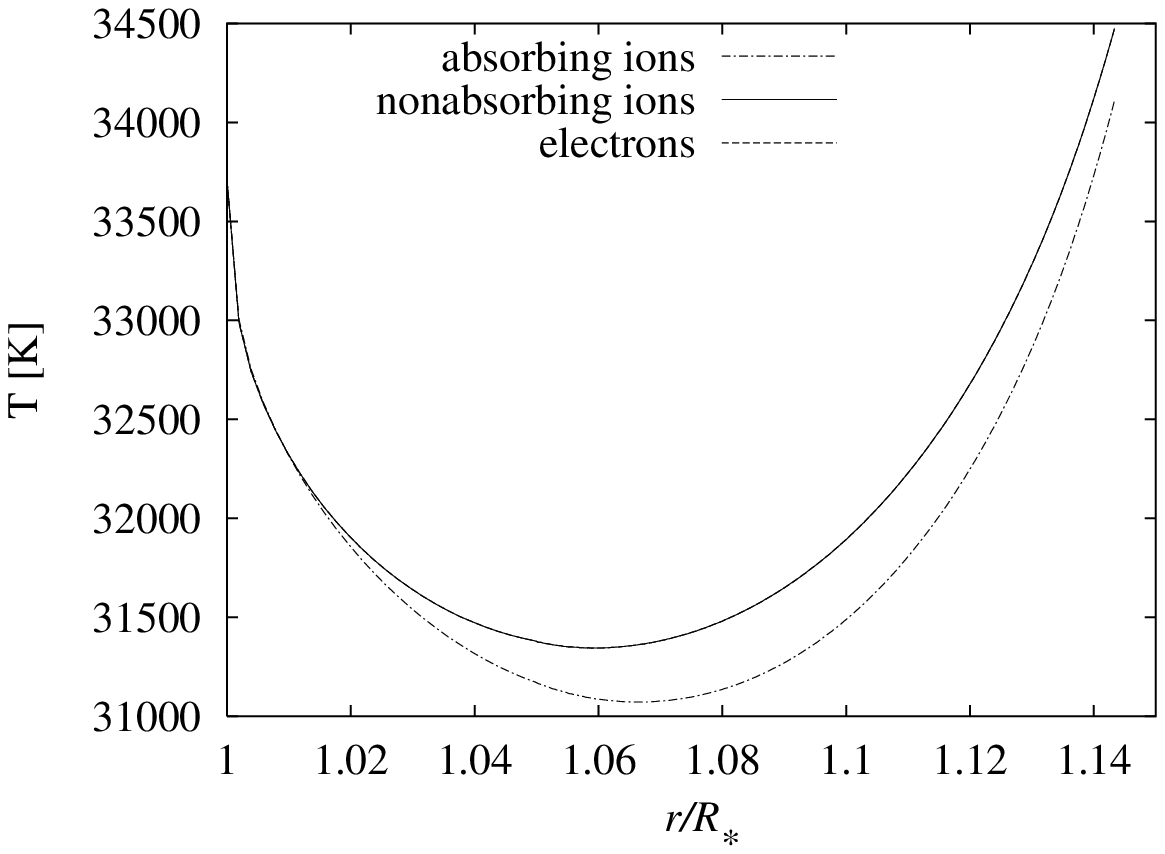}}
\caption[]{Calculated model of a star with $\log L/L_\odot=6.42$,
$T_\mathrm{eff}=40\,000\,\mathrm{K}$ and $Z/Z_\odot=0.001$. In the outer parts
of the wind the velocity difference is comparable to the sound speed,
$(v_{r\mathrm{i}}-v_{r\mathrm{p}})/\sqrt{\frac{2kT}{m_{\mathrm{p}}}}\approx 1$
and runaway instability occurs. Thus, the outer model boundary is
below this area. The GO 
heating has a negative sign
near the stellar surface and positive in the outer wind regions. Due to this
variation the temperature of absorbing ions is lower than the temperature of
other components.}
\label{stab}
\end{figure}

One of the models where such instability occurs is given in
Fig.\ref{stab}.
Note that the outer model boundary is selected in such a way that the possible
region of runaway instability is downstream.

\subsection{Possible fallback of nonabsorbing component}

For stars with even lower density, decoupling occurs in the region 
where the velocity of
the nonabsorbing component is lower than the 
{\jcb escape}
 velocity.
This causes either
the fallback of hydrogen and helium onto the stellar surface or possible
creation of some kind of clouds above the stellar surface (similar to
that suggested by Porter \& Skouza \cite{reac}). Note that, in this
case, time-dependent modeling is necessary to calculate the proper
structure of a multicomponent stellar wind.

\begin{figure}
\resizebox{\hsize}{!}{\includegraphics{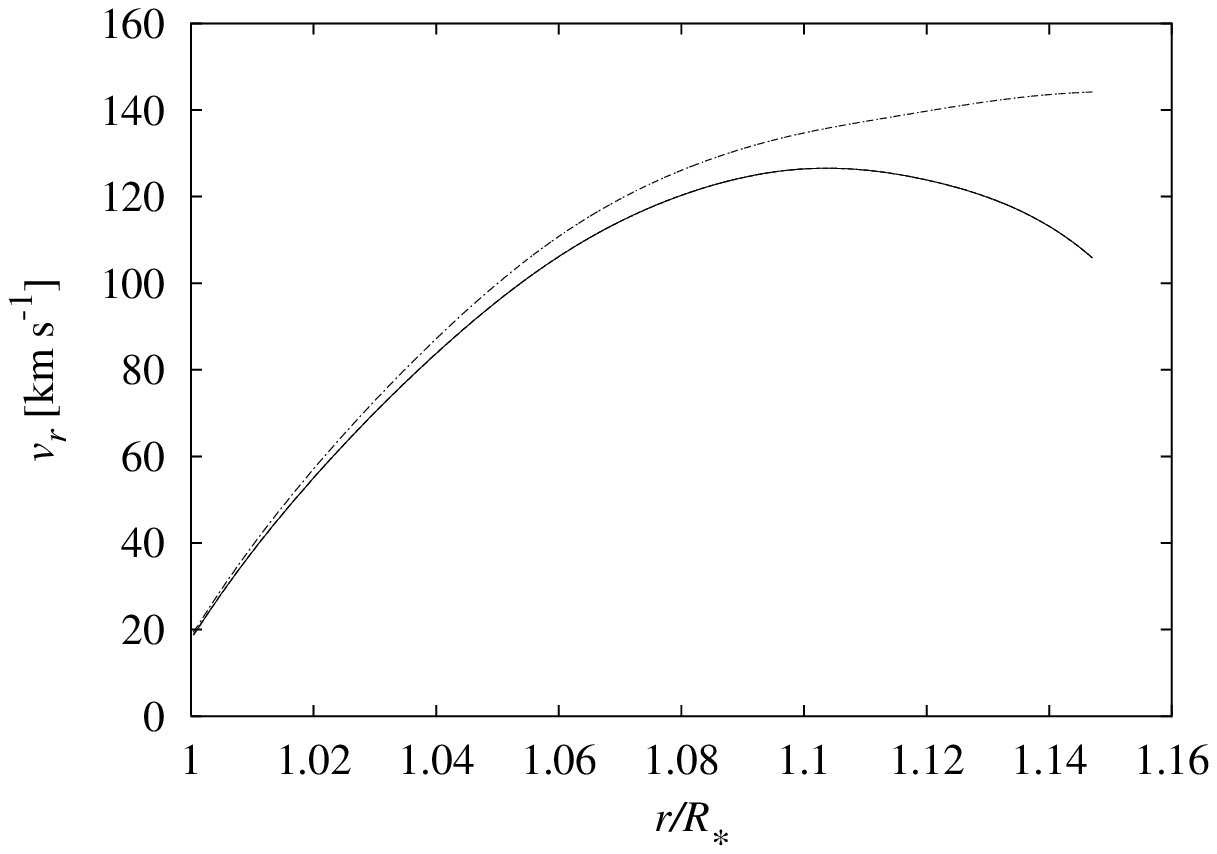}}
\resizebox{\hsize}{!}{\includegraphics{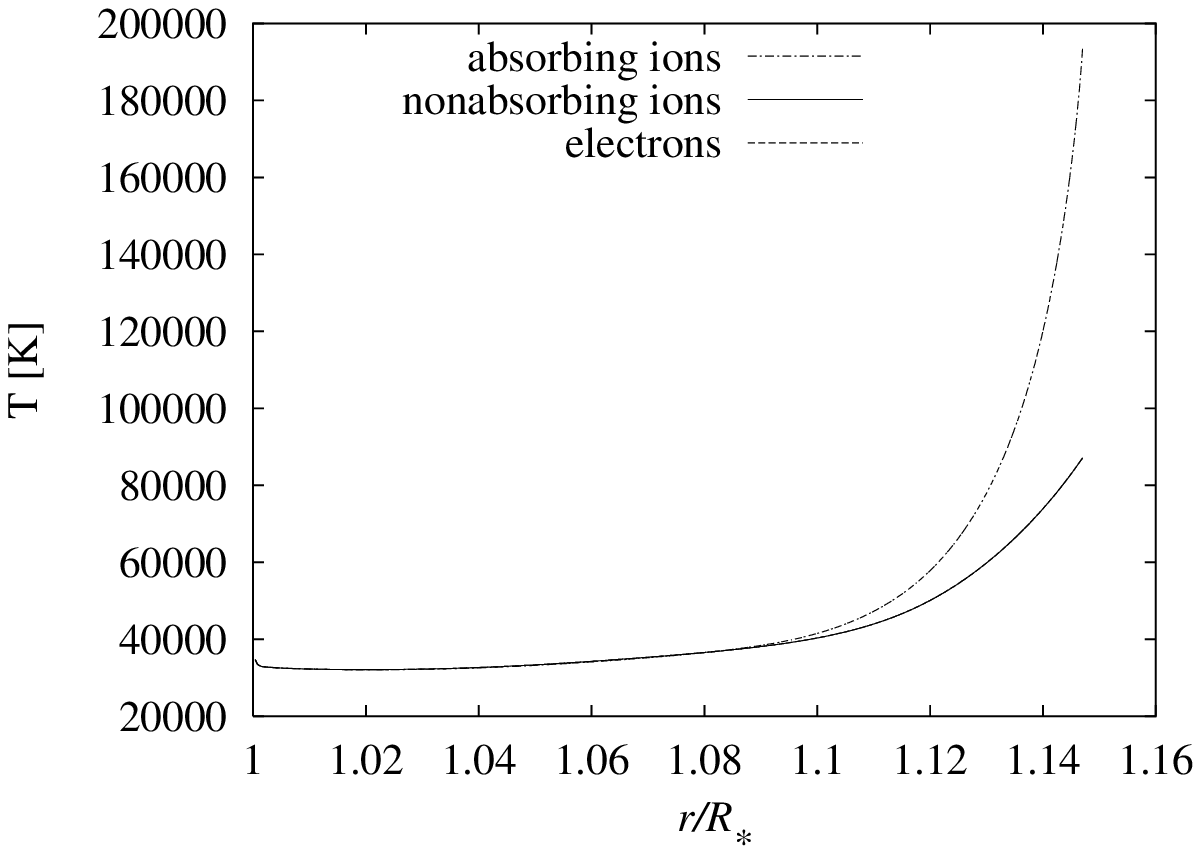}}
\caption[]{Calculated model of a star with $\log L/L_\odot=6.30$,
$T_\mathrm{eff}=40\,000\,\mathrm{K}$ and $Z/Z_\odot=0.001$. For this star the
absorbing component is not able to accelerate the nonabsorbing 
component and the
nonabsorbing component either form clouds around the star or falls 
back to the stellar
surface. The increase of temperature is caused by the
frictional and GO heating. Note that velocities and densities of
nonabsorbing component and electrons are nearly the same.}
\label{sup}
\end{figure}

These conclusions are demonstrated in Fig.\ref{sup}.
The absorbing component dynamically decouples from the nonabsorbing
component, accompanied even by decoupling of the temperatures.

\subsection{Formula for the maximal velocity difference}

{\jcb
\begin{figure}
\resizebox{\hsize}{!}{\input{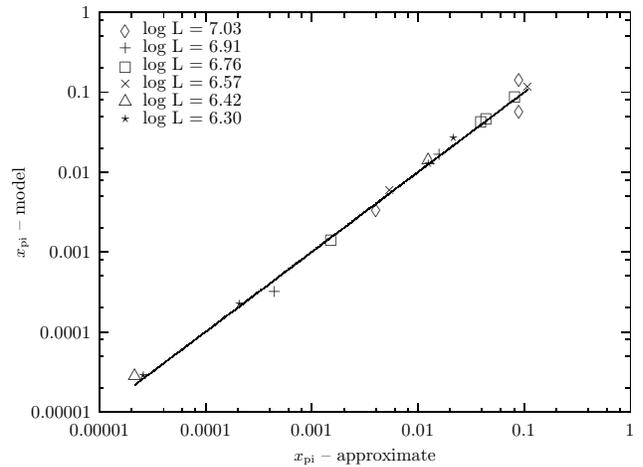}}
\caption[]{Comparison of the maximal velocity difference calculated with
approximate formula
\eqref{xpi}
and model ones.}
\label{porov}
\end{figure}

To test the reliability of our approximate formula for the maximal velocity
difference
\eqref{xpi}
we compared the approximate  velocity differences with
model ones for stars with negligible multicomponent effects. The results are
given in Fig.\ref{porov}. Clearly, formula
\eqref{xpi}
gives faithful
predictions for the maximal drift velocity between absorbing and nonabsorbing wind
components.

}

\section{Conclusions}

For most of the stellar  wind models
(moderate metallicity) calculated by Kudritzki (\cite{kudmet}) we did 
not find any significant multicomponent
effects, our calculations being in good agreement with his. 
Thus, radiatively driven winds of these stars can
be adequately  described by one-component models as done by Kudritzki.
However, for some low metallicity stars we found that 
the multicomponent wind nature
has a large effect
and 
must be included to describe 
stellar winds of such stars.
The importance of multicomponent effects increases
with decreasing metallicity and for any given star we are able to find
a metallicity value below which the multicomponent effects 
significantly influence the wind
structure. We found several consequences of multicomponent effects, 
ranging from
modified temperature structure to  runaway instability and possible fallback
of the nonabsorbing component. 
{\refi Compared to results obtained by KKII, the frictional and GO
heating are not so effective mainly due to the larger radiative cooling of
higher density winds.}
Note that the changed wind structure
influences the stellar radiative flux and may be specially
important in the UV or X-ray region.
Finally, due to their different terminal
velocities these stars cannot be used for
stellar
distance measurements using the wind
momentum-luminosity
relation
(cf. Kudritzki et al. \cite{kudwml}).

{\refi The validity of models presented here is influenced by some
simplifications used. Although they do not influence the basic results, the
real picture of multicomponent flows may be slightly different. Firstly,}
the radiative force is influenced by the temperature due to
the temperature dependence of ionization and excitation. Although in the case of
line-driven winds the occupation numbers are mainly given by the radiative
processes, the temperature may be important  especially in the case when large
frictional heating occurs. For precise calculation of the radiative force it
would be necessary to solve NLTE rate equations in the stellar wind.
%
{\refi Secondly, the radiative heating/cooling term may be influenced by complex
ionization and recombination processes}.
Finally, for
extremely-low density stellar winds there may be an insufficient amount of
{\refi metallic}
optically thick lines and H, He lines may become important.
{\refi We plan to adress these interesting issues using our NLTE wind code.
First results obtained using this code for normal stars are promising and will
be published elsewhere (Krti\v{c}ka \& Kub\' at 
2003).}

\begin{acknowledgements}
We are grateful to Dr. R. P. Kudritzki for discussion of his models.
{\jcb This research has made use of NASA's Astrophysics Data System.}
This work was supported by a PPARC Rolling Grant and
{\jcb by grants GA \v{C}R 205/01/0656 and 
205/02/0445, and by projects K2043105 and Z1003909.}
\end{acknowledgements}

\end{document}